# Structural short - range order and electronic properties of graphene at low temperatures

Bobenko N.G.[1,2], Egorushkin V.E.[1], Melnikova N.V.[3], Ponomarev A.N.[1,2]

1 Institute of Strength Physics and Materials Science of SB RAS, 2/4 Akademichesky Avenue, Tomsk 634021, Russia
2 National Research Tomsk Polytechnic University, 30 Lenin Avenue, 634050 Tomsk, Russia
3 V.D. Kuznetsov Siberian Physical Technical Institute of Tomsk State University, 1 Novosobornaja Square, Tomsk 634050, Russia

E-mail: phdmelnikova@gmail.com

**Abstract.**
A theoretical study of the DOS, resistivity and thermal conductivity of a metallized epitaxial graphene with impurities and structural inhomogeneous of the short-range order type is performed. The conditions for the appearance and disappearance of a gap in DOS are obtained for the case where the location of defects changes. A temperature dependence of relaxation time, DOS, resistivity and thermal conductivity is described only in the presence of the structures forming large-area 'cells' in the graphene layer.

**Introduction.**
The results of experimental investigations of graphene properties conducted in the recent decades [1-19] have caused a real 'graphene boom' and even led to creation of a separate area of nanotechnology dealing with graphene and other graphene-base two-dimensional materials. It was found that graphene has high carrier mobility, thermal and electrical conductivity; all these properties depend on the external influence on graphene and modification of its structure. In particular, it was observed that an introduction of different functional groups into the graphene plane not only alters its electronic conductivity but also provides a selective affinity for specific molecules from the external environment, including the biological ones [1]. Graphene properties also vary due to the substitution of the carbon atoms for the atoms of other chemical elements, in particular silicon or germanium. It was found that metal nanoparticles are selectively adsorbed on the edges of graphene defects, whereby their contours 'are drawn' by the chains of metallic nanoparticles that are clearly visible in the electron microscope. These defects are not chaotically arranged but form ordered structures on the carbon surface; one square micrometer of carbon surface may contain up to two thousand defects (reactive centers). Moreover, it was observed that graphene is synthesized only in the form of clusters [1-3].
In the experimental studies [4-6] of large-area graphene films, formed on the metal or semiconductor substrates, it was observed that individual carbon atoms create graphene islands around themselves, which coalesce into a continuous film on the substrate. The mismatch of the lattice orientations of these interlocking islands leads to the formation of grain boundaries determining the electronic properties of the graphene sheet.

A number of theoretical investigations [7, 8] showed that electronic properties of graphene with an ideal structure are largely determined by the space orientation of graphene, the edge (which is an analogue of chirality of carbon nanotubes (CNTs)), the sheet size, and by the imperfection of graphene structures. An interaction of graphene with the substrate is observed to create a band gap in the graphene electronic structure, so the graphene sheet symmetry is retained for at least one layer [4, 9]. This explains why an epitaxial graphene may have the same electronic properties as an isolated graphene sheet. The experimental data [9-11] confirm the change of the gap width in the electronic density of states (DOS) of the epitaxial graphene from zero to the forbidden zone width of the substrate, depending on its type.

The Schottky barrier between the metallic graphene layer and the semiconductor substrate is due to the doping of graphene. The properties of this barrier depend on the structure and passivation of the interface (introduction of graphene into the substrate structure). In addition, gases can dope graphene forming various impurity configurations (substitution, pyridine, pyrrole, and other types). Each of them can give rise to various changes in the graphene electronic properties [3].

It was shown theoretically [12] that the pyridine and pyrrole configurations did not lead to significant doping. Substitutional nitrogen has the most pronounced doping effect among the various types of impurities. Furthermore, in the case of substitution of atoms the crystal structure is not disturbed and no vacancies are formed. Thereby the high charge carrier mobility, which is so critical in electronic devices, retains [13]. The process of transformation of one type of impurity into some other is accompanied by filling the electronic states of the conduction band of graphene and its metallization.

The experimental investigations of the doped graphene [14-17] showed that annealing of graphene with about 2 at.% nitrogen, with the pyridine nitrogen prevailing, results in transformation of the most part of the latter into substitutional nitrogen. It was also found that adsorption of hydrogen atoms on graphene surfaces leads to the formation of a band gap in the electronic structure, and the gap width can be controlled.

It is well-known that there is no gap in the DOS when a single hydrogen atom is adsorbed on the surface, regardless of the place of its adsorption [14]. The difference in the relative position of two hydrogen atoms on the surface of graphene results in a change in the gap width of the DOS from 0.6 to 3.1 eV. In the case of multi-layer coatings, the electrons are transferred from the graphene plane to the layer of hydrogen adatoms [17], which also leads to the gap opening. In the structure of the grafan type, where hydrogen atoms sit on the graphene surface on both sides alternately, the gap is also formed [15]. The authors of [15] noted that the present-day challenge is to describe the correlation between the width of the gap that is formed during hydrogen adsorption and the value of the hydrogen adatom concentration and position on the surface.

The change in the graphene structure influences not only the DOS, but also the graphene electronic properties. In [18], an addition of nitrophenyl groups was shown to change the electronic structure and transport properties of epitaxial graphene, so that conductivity changed from the metallic to semiconductor type.

In [19], the electron transfer in graphene was experimentally investigated and a completely different behavior of electrical resistance was observed for two different densities of charge carriers: for n = 0 (Dirac point) and

$3\times10^{12}$ cm$^{-2}$ (high density). Given a high density of the carriers, there is a noticeable monotonic decrease in the electrical resistance normalized to that at room temperature ρ(T)/ρ(300K), which is observed with an increase in temperature. The authors of [19] attributed this phenomenon to the influence of electron-phonon scattering appearing in the scattering mechanism in the limit of high conductivity densities. A monotonic increase in ρ (T)/ρ (300K) with an increase in temperature may indicate a nonmetallic behavior of the Dirac point conduction.

Thus, it is extremely important to investigate the influence of the defect structure, type and location of the adsorbed gases in grapheme structure, as well as the type of substrate on the formation and width of the gap in DOS of epitaxial graphene. Moreover, it is necessary to take into account that at low temperatures the rearrangement of atoms forming the short-ranged ordered domains, accompanied by creation of new chemical bonds, may take place in graphene. In order to understand the electron transfer mechanisms in graphene it is necessary to study its DOS and resistivity.

In the present paper, we perform a theoretical study of the DOS and resistivity of a metallized epitaxial graphene with impurities and structural inhomogeneous of the short-range order type. The conditions for the appearance and disappearance of a gap in DOS are obtained for the case where the location of defects changes. The temperature dependence of electrical resistance of graphene on a substrate is also described. Our approach relies on the assumption that graphene contains short-range order structures.

**Model**.

For the qualitative and quantitative description of defects in the graphene structure, we are going to use the concept of short-range order parameter [20]. In the defect-free structure, the short range order parameter is taken to be equal to 0. If the first coordination sphere contains topological defects or the second or subsequent coordination spheres contain atoms of a different kind, the order parameter is positive. When in the first coordination sphere there are atoms different from carbon, this parameter is negative. Figure 1 shows an example presenting different locations of defect atoms in the graphene structure.

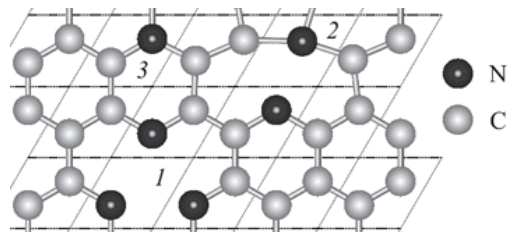

**Fig.1.** The main types of nitrogen impurities in graphene: pyridine nitrogen (*1*), pyrrole nitrogen (*2*), and substitution graphite-like nitrogen (*3*) [3].

The low-temperature characteristics of the graphene DOS will be described using the method of the temperature Green functions [21], taking into account the elastic electron scattering on impurities and structural inhomogeneities of the short-range order type. As we demonstrated for carbon nanotubes this theory [22-25] and experimental data [26-28] are in good agreement.

In order to describe graphenes as systems with impurities and local structures with the new short-range order, we shall introduce a random field of the adsorbed one-kind atoms (for simplicity)

$$V(\vec{R}) = \sum_i c(\vec{R}_i) U(\vec{R}_i - \vec{R}), \qquad (1)$$

where $c(\vec{R}_i) = c + \delta c(\vec{R}_i)$ are the filling numbers, $\delta c(\vec{R}_i)$ are fluctuations of concentration in site $\vec{R}_i$, $c$ is the averaged (macro-) concentration, finally, $U$ is the site potential of an electron in graphene. The Fourier image $\langle \delta c(\vec{R}_i) \delta c(\vec{R}_j) \rangle \sim \langle |c_k|^2 \rangle$ determines the short-range order structure, so $\langle |c_k|^2 \rangle = \frac{c(1-c)}{N} \sum_{i=1}^{N} \alpha_i \cos k \cdot \vec{R}_i$ [20].

Here $\alpha_i$ are the short-range order parameter and $N$ is the number of atoms inside the structure inhomogeneity of the short-range order type. The angular brackets mean averaging over a random field.

For all $\alpha_{i \neq 0} = 0$ and $c \to 0$, $\langle |c_K|^2 \rangle = \frac{c(1-c)}{N}$ determines the input system impurities. At $\alpha < 0$, the 'foreign' atoms are located in the first coordination sphere, and ordered structures are formed. At $\alpha > 0$, these atoms are arranged in the second coordination sphere, leading to stratification of the local range of graphene.

Recall now that our principal assumption is the presence of short-range order structures formed in the graphene layer during its synthesis and subsequent adsorption of gases on its surface. The short-range order parameter $\alpha$ is the quantitative and qualitative characteristics of structure defects in the proposed model [20]. In this work, we calculated $\alpha$ for different configurations of foreign atoms for the cases of ordering (Figure 2a) and stratification (Figure 2b), and for the substitutional, pyrrole and pyridine structures in a two-dimensional graphene layer. The results of our calculation are given in Table 1.

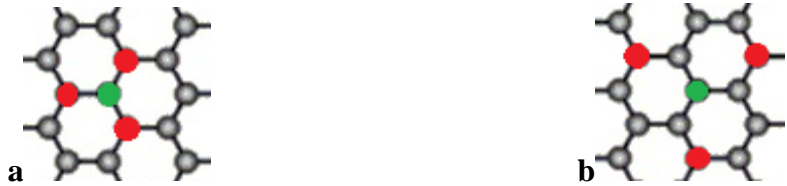

**Fig. 2.** An illustration of the case of ordering (a) and stratification (b) of atoms in graphene: carbon atoms (gray circles), carbon atom considered as centers of a short-range ordered domain (green circles) and impurity atoms (red circles).

| Short-range order parameter $\alpha_i$ <br> Type of defect structure | $\alpha_1$ | $\alpha_2$ |
|---|---|---|
| Ordering | -0,17 | - |
| Stratification | - | 0,64 |
| Substitution | 0,46 | - |
| Pyrrole | 0,50 | 0,16 |
| Pyridine | 0,11 | - |

**Table 1.** The values of short-range order parameter for the most common configurations of structural defects in the two-dimensional graphene in the first and second coordination spheres.

It is evident from Table 1 that the values of short-range order parameters for the cases of ordering and stratification have different signs. All other structures correspond to the case of stratification. The pyridine structure has the lowest short-range order parameter. For a more energetically favorable structure of the substitution type, parameter $\alpha_1$ is equal to 0.5 in the first coordination sphere and in the case of the pyrrole structure, while in the second coordination sphere $\alpha_2$ differs from zero for both types of defects.

To calculate the single-particle properties, such as electron relaxation time and the DOS, let us present the single-particle Green function as follows:

$$G = G_0 + \sum_i G_i^{(1)} + \sum_{ij} G_{ij}^{(2)} + ..., \qquad (2)$$

where $G_0$ is the electron Green function in a pure graphene and $G_i^{(1)} = c(R_i)\int G_0(\vec{r}, R)U(R-R_i)G_0(R,\vec{r}')dR$. The detailed expression for $G_{ij}^{(2)}$ is given in [29]. For simplicity, we have assumed that $U(R-R_i) = U_0\delta(R-R_i)$, because the radius of its action is small in comparison with the distance between the atoms (disperse system).

In a $p$ – representation, after averaging over disorder we have

$$\langle G \rangle = G_0 + G_0^2 \Sigma, \qquad (3)$$

where $\Sigma(\vec{p},\varepsilon) = -i2\pi C U_0^2 \nu_0 \left(1 + \langle |c(\vec{p})|^2 \rangle\right)$ is the self-energy part including elastic electron scattering on impurities and short-range order structural inhomogeneities. Averaging $\langle |c(\vec{p})|^2 \rangle$ over the $\vec{p}$ vector angle, we obtain the following expression for $\Sigma$:

$$\Sigma(\varepsilon) = -i2\pi c U_0^2 \nu_0 sign\varepsilon \left[1 - \frac{1-c}{N}\alpha_i\left(1 + \frac{R_i^2 m}{\hbar^2}(\varepsilon + \varepsilon_F + i0 sign\varepsilon)\right)\right]. \qquad (4)$$

Here $\nu_0 = \dfrac{m}{2\pi\hbar^2}$ is the original DOS at the Fermi level in an ideal graphene (without any defects).

The first term in (4) corresponds to the scattering on impurities and the second one describes the elastic scattering of electrons on structural inhomogeneities of the short-range order type.

To calculate the relaxation time $\dfrac{1}{\tau} = -\text{Im}\Sigma$ and the contribution to DOS $\Delta\nu = -\dfrac{1}{\pi}\text{Im} Sp(\langle G\rangle - G_0)$, in $G_0 = \dfrac{1}{\varepsilon - \varepsilon_p + i0}$ we shall use the dispersion relation $\varepsilon_p = \hbar k \upsilon_F$ for a perfect graphene plane near the Fermi level $\varepsilon_F$, where $\upsilon_F \approx 10^6$ m/s [30].

Using (4) and going from energy to temperature, we arrive at

$$\frac{1}{2\tau} = \frac{1}{2\tau_{imp}}\left[1 + \frac{1-c}{N}\alpha(\beta T - 1)\right], \qquad (5)$$

where $\beta = \pi \dfrac{R^2 m}{\hbar^2} k$, $\dfrac{1}{\tau_{imp}} = \dfrac{2\pi}{\hbar} c U_0^2 \nu_0$.

From (5) it is clear that the inverse relaxation time depends on temperature, concentration of defects and short-range order parameter values. The first term in (5) corresponds to the scattering on impurities but it is independent of both $T$ and $\alpha$. The second term describes multiple elastic scattering of electrons on a new structure of short-range order and depends of $\alpha$ and $T$.

For 3-dimensional structures with the short-range order, $\beta = \dfrac{2\sqrt{2\pi}(1-c)m^{3/2}}{v_0 N}$ [23] and it does not depend on the size of the short-range ordered domain $R_i$. In 2-dimensional graphene structures, $\beta$ strongly depends on $R_i$. If the coordination sphere radius is determined by the ideal unit cell, $R \approx 0,25 nm$, then $\beta \sim 10^{-3} K^{-1}$ and the relaxation time does not depend on temperature in the temperature interval considered. As a result, the main mechanism of electron scattering is due to impurities.

Real graphene is in fact formed by unit cells with $R \approx 2,5 nm$, then $\beta \sim 10^{-1} K^{-1}$. For this case, the temperature dependence $\tau(T)$ is presented in Figure 3 for the structures with ordering and stratification.

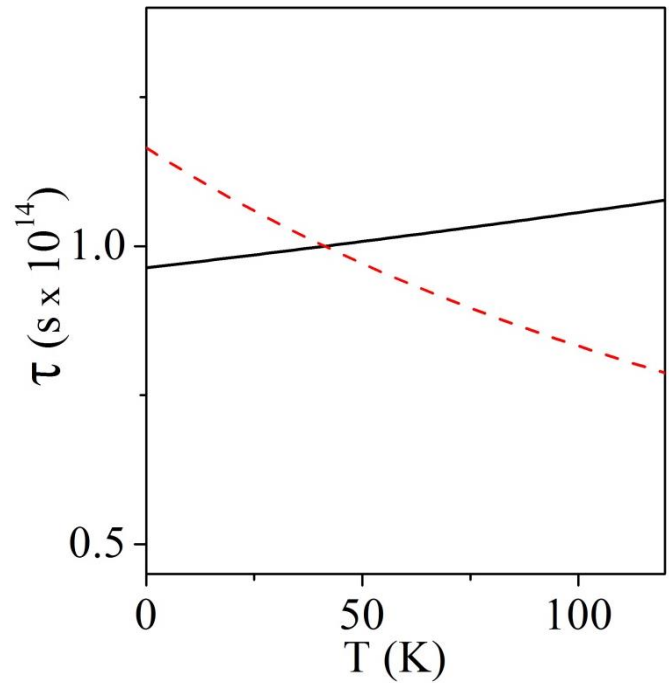

**Fig. 3.** The temperature dependence of relaxation time for the cases of ordering (solid line, $\alpha$=-0.17) and stratification (dashed line, $\alpha$=0.64) of graphene.

It is seen from Figure 3 that the relaxation time increases with the temperature for the case of ordering and decreases for the case of stratification. At $T\sim 100K$, these relaxation times are different by 30%. As a result, the mean free path ($L=v_F\tau$) shows a different dependence on temperature for the cases of graphene ordering and stratification. At $T<40$ K (the temperature at the intersection point of the $\tau(T)$ curves for the cases of ordering and stratification, the mean free path is larger for stratification structures and smaller for the ordering type structures. This corresponds to a situation, where stratification electrons scatter on the second coordination sphere, while in the cases of ordering they scatter on the first coordination sphere.

With increasing temperature ($T>40$ K), the opposite situation takes place: the mean free path length is larger for the ordering than that for the stratification type structures. Apparently, at the temperatures above 40 K the electron scattering on phonons is more meaningful. In this regard, we would like to elaborate on [31], where an experimental study of the nonlinear nature of electrical conduction in monolayer graphene devices on silica substrates was performed and a nonmonotonic dependence of the differential resistance on applied dc voltage bias across the sample was observed. At the temperatures below ~70 K, the differential resistance was found to exhibit a peak near zero bias, which can be attributed to self-heating of the charge carriers. The authors of [31] estimated such characteristics as the electron-phonon relaxation length, the corresponding relaxation time, and the diffusion coefficient. Table 2 summarizes their estimations, the data from [3] and the results of our calculations of relaxation time, mean free path, and diffusion coefficient. One may see that all values are of the same order of magnitude, which indicates the importance of electron scattering on the short-range ordered domains. At low temperatures, this scattering may be more important than electron scattering on phonons.

| Quantity | Estimations for electron-phonon relaxation [31] | Data from [3] | Calculations at $T<100$ K [this work] |
|---|---|---|---|
| Relaxation time $\tau$ | $0.36$–$1.08 \cdot 10^{-14}$ s | $10^{-14}$–$10^{-12}$ s | ~$10^{-14}$ s |
| Mean free path length $L = v_F \tau$ | 40–100 nm | 10–100 nm | ~10 nm |
| Diffusion coefficient $D = v_F^2 \tau / 2$ | 18–50 cm$^2$/s | 50–5000 cm$^2$/s | ~50 cm$^2$/s |

**Table 2.** The data from [31] and [3] and our calculations of relaxation time, mean free path length and diffusion coefficient.

Now let us calculate the contribution to the DOS $\Delta \nu$ due to electron scattering on impurities and short-range order structures. The Green function averaged over disorder is given by

$$\langle G \rangle = \frac{1}{\varepsilon - E_p + \frac{i}{2\tau} Sgn\varepsilon}, \qquad (6)$$

where $E_p$ is the energy spectrum of a 'dirty' graphene (graphene with defects). We may assume that the energy spectra of an ideal graphene and a dirty graphene differ negligibly, i.e. $\varepsilon_p \approx E_p$. Then we obtain

$$\langle G \rangle - G_0 = \frac{\frac{i}{2\tau}}{\left(\varepsilon - \varepsilon_p + \frac{i}{2\tau}\right)(\varepsilon - \varepsilon_p)}.$$

As a result,

$$\Delta \nu = \frac{1}{2\tau} \int \frac{p \, dp}{(\varepsilon - \varepsilon_p)^2 + \left(\frac{1}{2\tau}\right)^2}. \qquad (7)$$

Integrating (7) over the momentum, we obtain the final expression for the contribution to the DOS near $\varepsilon_F$ in graphene with impurities and structural short-range order defects given by

$$\Delta \nu = \frac{1}{\hbar^2 \upsilon_F^2}\left[\frac{\hbar}{2\tau}\ln\left(1+\frac{p_F\upsilon_F(p_F\upsilon_F - 2\varepsilon))}{\varepsilon^2 + \left(\hbar/2\tau\right)^2}\right) + 2\varepsilon \, arctg\left(\frac{p_F\upsilon_F\left(\hbar/2\tau\right)}{\left(\hbar/2\tau\right)^2 + \left(\varepsilon - p_F\upsilon_F\right)\varepsilon}\right)\right] \quad (8)$$

The contribution to the DOS (8) depends on temperature, defect concentration and short-range order type (inhomogeneous structure).

To calculate the resistance of the epitaxial graphene, we use the expression for electrical resistivity in disordered metallic systems with short-range order

$$\rho(T) = \frac{m}{e^2 n\tau} = \frac{m}{e^2 n\tau_{np}}\left(1 + \frac{1-c}{N}\alpha(\beta T - 1)\right), \quad (9)$$

where $e$ and $m$ are the electron charge and mass, respectively, $n$ is the number of current carriers, and $\tau$ is electron relaxation time. We use Wiedemann-Franz low for thermal conductivity.

**Results and discussions.**

Calculating the contribution to the DOS and resistivity, we restricted ourselves to the following parameters: the temperature was changed from 0.1 to 120 K, the defect concentration – from 0 to 0.1, and the short-range order parameters – from -0.17 to 0.64. The latter corresponds to the different short-range order types: when $\alpha$ is negative, defects are located in the second (first) coordination sphere and when $\alpha$ is positive - in the first (second) one [20]. Finally, $\nu_0$=0.1 eV$^{-1}$ and $U_0$=0.05 eV.

Equations (8) and (9) allow us to analyze how the change in the sign of short-range order parameters affects the contributions to the DOS at a fixed temperature near the Fermi level. It should be borne in mind that the sign of the short-range order parameter is changed from negative to positive during doping, annealing, or degassing of graphene.

The DOS, calculated for the cases of ordering and stratification of graphene using the temperature Green function method, is presented in Figure 4a. The gap opening at the Fermi level is due to the negative contribution to the DOS from electron scattering on structural inhomogeneities of the short-range order type. In the case of stratification, this contribution is positive and results in metallization of graphene. Thus our model allows describing both dielectrization and metallization of graphene.

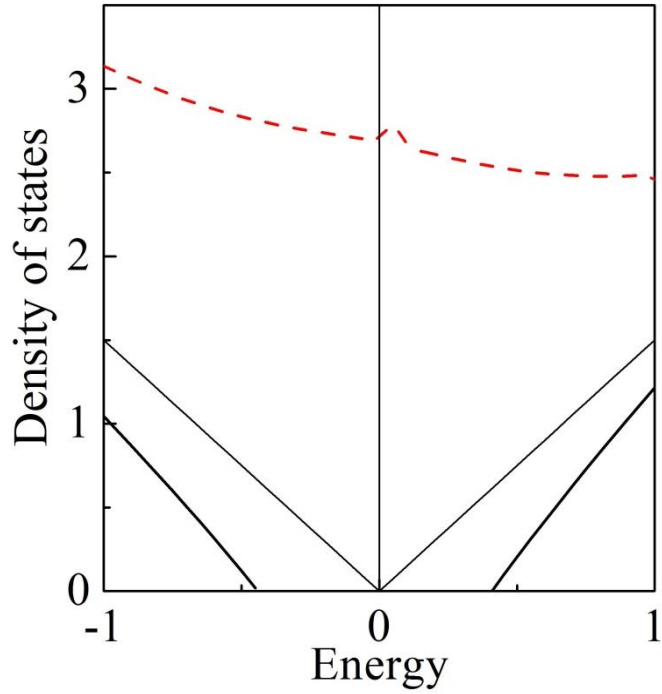

**Fig. 4.** The DOS calculated by the temperature Green function for the ideal graphene (dotted line), for $\alpha = -0.17$ (solid line) and 0.64 (dashed line) at T=5K (*a*).

We have also calculated the DOS by the density functional method. The results of this calculation are presented in [35]. Comparing Figure 4a and results [35], one may see that for an ideal graphene the DOS near the Fermi energy is linear in both methods of calculation and is consistent with other calculations [30]. However, the density functional method does not allow describing the DOS changes when the graphene is ordered (Figure 1a) or stratified (Figure 1b). This method takes into account the changes in the number of electrons per atom only, when foreign atoms are available; this results in a 1 eV shift of the Fermi level and a 15% increase in the DOS. In [32], the experimental data for the DOS in graphene with different geometric capacitance are presented. An increase in the geometric capacitance is found to increase the DOS at the Fermi level compared with the DOS calculated in the ideal graphene. In our opinion, this may be due to the defects present in the graphene structure, because the contribution to the DOS associated with the short-range order, also increases when the short-range order parameter rises at a fixed temperature.

Next, let us analyze the influence of the sign of the short-range order parameter and that of concentration and temperature on the contribution to electrical resistivity of graphene. Figure 5 presents the temperature dependence of electrical resistivity calculated at $\alpha = 0.64$ and $-0.17$ and the experimental electrical resistivity data [33]. One may see that with increasing temperature the contribution to resistivity either increases or decreases. In our model, this may be accounted for by the change of the sign of the short-range order parameter. Calculating the resistivity with the following parameters: $v_0 = 1{,}6*10^{-18} J^{-1}$, the square of a unit cell $s \sim 10^{-18} m^2$ and $\dfrac{m}{n} = \dfrac{3s}{v_0 v_F^2}$, we obtained its value equal to $f\rho \sim 10^3 \, \Omega$. This is well confirmed by the experimental data [33].

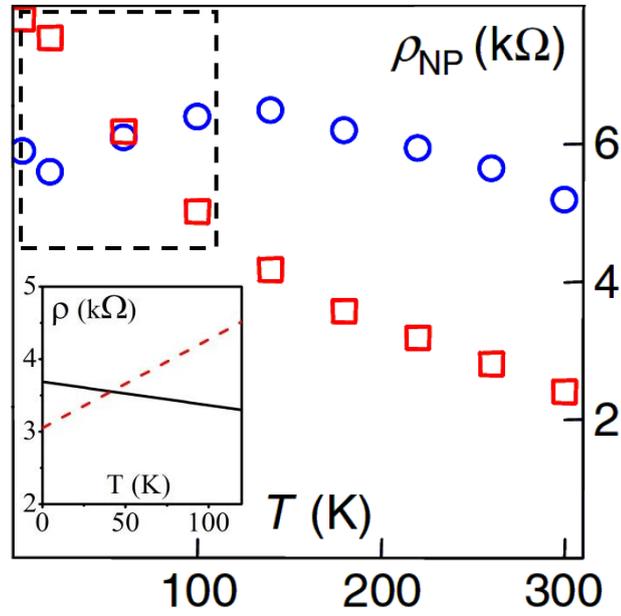

**Fig. 5.** The resistivity for the single-layer (circles) and double-layer (squares) graphene [30] and the temperature dependence of electrical resistivity calculated at $\alpha = 0.64$ (dashed line) and -0.17 (solid line) (insert).

In [31], the dependence of differential resistance on voltage at different temperature values (4.2, 17, 26, 35, 83, and 107 K) was investigated (Figure 6a). At $T<50K$ near the zero potential point there is a local maximum in $R(U)$ whose value decreases with increasing temperature. At $T> 50K$ this maximum disappears. We mentioned above that the authors [31] had attributed the temperature changes in this peak to self-heating of the charge carries caused by phonons. We assume that the origin of the maximum in $R(U)$ may be associated with the contribution from electron scattering on 'foreign' atoms into the DOS in non-ideal graphene. To check this assumption, we calculated the energy dependence of the differential resistance using the Einstein formula and the contributions to the DOS described by eq.(8) for $T= 4.2$ and 35K, $\alpha=0.64$. The calculated curves are presented in Figure 6b; it is evident that that the calculated differential resistance $R(U)$ also has maximum, whose value decreases with increasing temperature as in the experiment [32].

Thus, we may argue that foreign atoms in graphene [31] at $T<50K$ are located in the second coordination sphere; it is clear from Figure 6b that at T <50 K the contribution from electron scattering on the short-range order structures may be decisive.

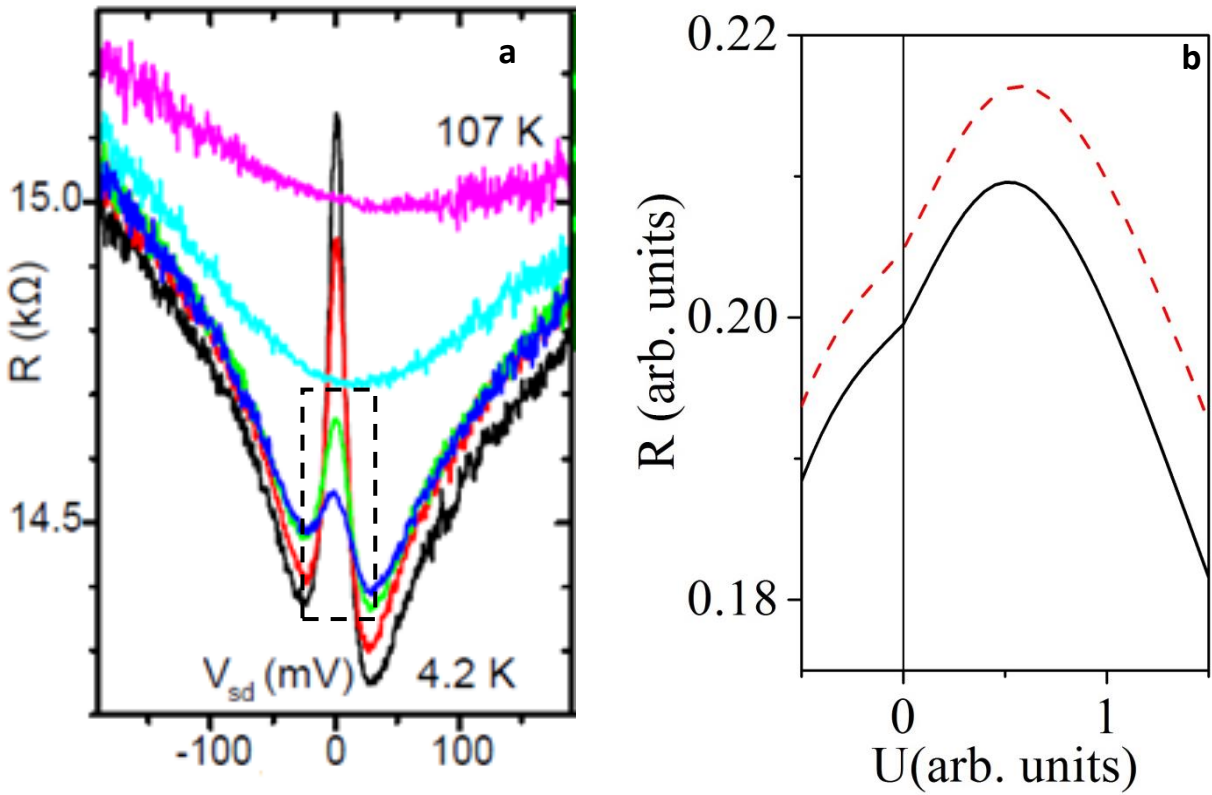

**Fig. 6.** The differential resistance as a function of DC source-sink bias measured at different bath temperatures: TB = 4.2, 17, 26, 35, 83, and 107 K from bottom to top [31] (a). The energy dependence of differential resistance *R(U)* calculated for *T*= 4.2 (solid line) and 35K (dotted line) and $\alpha$=0.64 (b).

Using the Wiedemann-Franz law and Equations (9) we have calculated heat conductivity and found that unlike the temperature dependence of electrical resistivity (Figure 5), the type of the temperature dependence of heat conductivity is the same for the cases of graphene ordering and stratification (Figure 7). The result of our calculation is in a good qualitative and quantitative agreement with the experimental data for heat conductivity [34] the authors of which also concluded that at low temperatures, the heat is transferred by electrons.

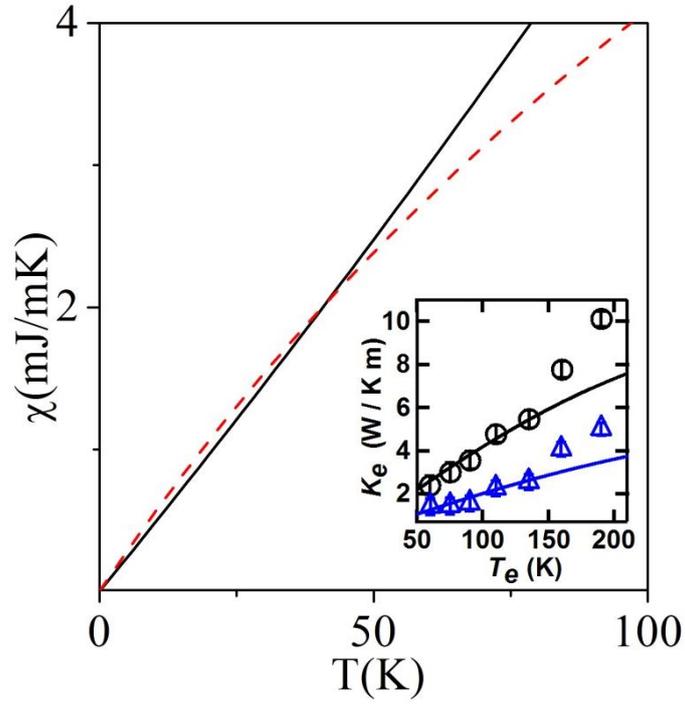

**Fig.7.** Temperature dependence of heat conductivity of graphene for the cases of ordering (solid line, α=-0.17) and stratification (dashed line, α=0.64). On insert: The electronic thermal conductivity, $K_e$ at $V_G$ = 5, 3, and 2 V corresponding to $n_{tot, T=0} \approx$ 1.1, 0.7, 0.5 × $10^{11}$ $cm^{-2}$ [34].

**Conclusions.**

In this paper we have shown that only in the presence of the structures forming large-area 'cells' in the graphene layer there is a temperature dependence of relaxation time, DOS and resistivity, which is sensitive to the concentration of defects and the types of short-range order (structural inhomogeneities). Adsorption of gases and deep graphene penetration into the substrate lead to the appearance of atoms of a different kind in the first coordination sphere. As it follows from our results, this case corresponds to the negative short-range order parameter value. The structures with this atomic arrangement are known as ordered ones. If the impurity or doped atoms are in the second coordination sphere, then the short-range order parameter is positive and graphene is stratified.

We have also shown that in the cases of graphene ordering or stratification the electron relaxation time behaves itself in a different way. It decreases in the former case, when the temperature rises, and decreases in the latter. The mean free paths and diffusion coefficients calculated here are of the same order of magnitude as estimated in [31] for electron-phonon scattering. That is why at low temperatures the electron scattering on the short-range order structures may be even more important than that on phonons. The same result has been confirmed with our calculation of thermal conductivity and [34].

In the case of stratification, the calculated contribution to the DOS is positive. This suggests that the presence of impurities or the doped atoms may be responsible for the metallization of graphene in the case where the gap at the Fermi level is closed. When the short-range order parameter is positive, the temperature behavior of resistivity is typical for metals [33]. In the ordered graphene, the contribution to the DOS from electron scattering on the short-range order type structures, changes its sign to the opposite one, which results in the opening of a gap in the DOS; the respective contribution to electrical resistivity decreases when the temperature rises. The depth of the minimum in the DOS depends on temperature, thus it increases with $T$ in a metallized graphene and decreases in a gas-saturated graphene, which is also validated by the experimental data [19].

**Acknowledgements**.

The research was sponsored by RFBR (project number 16-32-00398 mol_a), Marie Curie International Research Staff Exchange Scheme Fellowship within the 7th European Community Framework Program (grant agreement number 612552) and Program of Fundamental Scientific Research of State Academies of Sciences for 2013–2020 (project number 23.1.2).